\documentclass[11pt, a4paper]{article}
\usepackage[a4paper, bottom=3.0cm, top=3.0cm, inner=2.5cm, outer=2.5cm]{geometry} 
\usepackage[dvipsnames]{xcolor}
\usepackage[inkscapelatex=false]{svg}
\usepackage{graphicx}
\usepackage{float}
\usepackage{subcaption}

\usepackage{comment} 
\usepackage{chemformula} 
\newcommand{\InOx}[0]{\ch{In2O3}} 
\usepackage{setspace} 

\usepackage[colorlinks=true, citecolor=blue, linkcolor=blue, urlcolor=blue]{hyperref}

{ \begin{itemize}
    \setlength{\itemsep}{0pt}
    \setlength{\parskip}{0pt}
    \setlength{\parsep}{0pt}}
{ \end{itemize}} 

\begin{document}

\setstretch{1.25} 

\begin{center}
\textbf{\Large The absence of a central metal ion destabilizes phthalocyanine on \ch{In2O3}(111)}\\[0.5cm]

Viktoria Waidbacher$^1$, Sarah Tobisch$^1$, Faith J.\ Lewis$^1$, Moritz M.J.\ Eder$^1$, Michael Schmid$^1$, Gareth Parkinson$^1$, Ulrike Diebold$^1$, Margareta Wagner$^{1*}$ 
\\[0.5cm]

$^1$Institute of Applied Physics, TU Wien, 1040 Vienna, Austria\\
$^*$Corresponding author: wagner@iap.tuwien.ac.at
\end{center}

\setstretch{1.5} 

\subsection*{Abstract}
Metal phthalocyanines (MPc) are a versatile molecular platform for applications ranging from organic optoelectronic devices to single-atom catalysis (SAC). Their adsorption and layer formation on the prototypical transparent electrode substrates of organic optoelectronic devices is directly relevant for charge injection and transport across the organic--oxide interface. Moreover, the well-defined M--N$_4$ coordination of the metal cation defines their activity as SACs for (electro-)catalysis. Here, the adsorption of the metal-free phthalocyanine (H$_2$Pc) is characterized on \ch{In2O3}(111) using low-temperature STM, nc-AFM, and STS. \ch{In2O3} is not only a model system of indium tin oxide (ITO) but also an active catalytic material for CO$_2$ reduction. H$_2$Pc adsorbs in the same site and configuration reported for copper phthalocyanine [J.\ Mater.\ Chem.\ C 13, 17650--17661 (2025)] and for the majority of cobalt phthalocyanine [Surf.\ Sci.\ 722, 122065 (2022)]. Despite this shared preference in adsorption site, H$_2$Pc cannot be organized into extended and ordered monolayer structures by gentle annealing: the molecule starts to decompose at $\approx$50\,$^\circ$C, well below the temperature used to grow monolayers of CoPc and CuPc. Self-metalation is not observed on stoichiometric \ch{In2O3}(111). On the reduced surface where In$^0$ adatoms are present, new H$_2$Pc-related features appear but cannot be identified by imaging only. 
The comparison of H$_2$Pc with CoPc and CuPc identifies distinct roles of the central metal ion in the metal-Pc/\ch{In2O3}(111) systems: it acts as a structural anchor that stabilizes the macrocycle against decomposition on the surface, and it modifies the frontier-orbital character in ways that determine whether a second adsorption configuration is populated.

\clearpage
\section{Introduction}

\begin{figure}[H]
    \centering
    \includegraphics[width=0.8\linewidth]{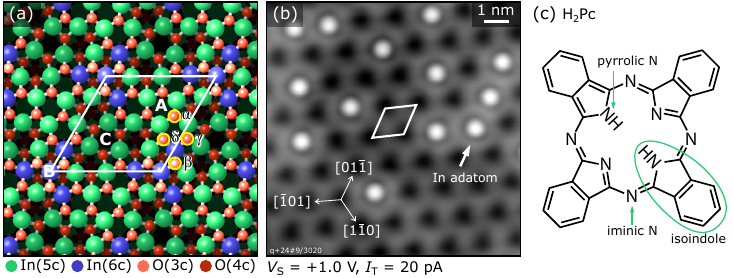} 
    \caption{\InOx(111) and H$_2$Pc. (a) Atomic model of \InOx(111) introducing the unit cell as used in this work, its three high-symmetry axes (A, B, and C), and the four inequivalent surface O anions O($\alpha$--$\delta$). In the top right corner is an In adatom in B site, characteristic for reduced \InOx(111). (b) Empty-states STM image of slightly reduced \InOx(111) featuring In adatoms in B. In the areas without adatoms, the four In(6c) cations around symmetry point B together as dark triangular features due to their low density of states.~\cite{MW2014}. (c) Structure of H$_2$Pc identifying the isoindole building block as well as chemically different N atoms.}
    \label{fig:InOx_surf}
\end{figure}

Phthalocyanines (Pcs) are among the most widely studied model systems in molecular surface science~\cite{GOTTFRIED}. Their planar $\pi$-conjugated macrocycle, built from four isoindole units linked by iminic ($=$N--) nitrogen atoms (Fig.~\ref{fig:InOx_surf}), either hosts two pyrrolic hydrogen atoms or a single (transition) metal ion M$^{2+}$ in a fourfold-symmetric M--N$_4$ coordination environment. Varying the central transition metal atom (Mn, Fe, Co, Ni, Cu, Zn) systematically tunes spin state, frontier-orbital character, and redox behavior while leaving the molecular framework essentially unchanged. The concept of an organic ligand with a rigid, chemically tunable single active site, has made metal-phthalocyanines (MPcs) a long-standing platform for applications ranging from pigments to organic optoelectronic devices, and for fundamental studies of molecular magnetism, single-molecule spintronics, and photo- and electrochemical catalysis~\cite{PC_intro_qubits, PC_intro_spin1, PC_intro_spin2, PC_intro_catal1, PC_intro_catal2}. Beyond these established uses, the atomically defined square-planar M--N$_4$ coordination of MPcs has increasingly attracted interest in the context of single-atom catalysis (SAC): each MPc molecule delivers a chemically well-defined, aggregation-resistant M--N$_4$ active site directly to a substrate, without the structural heterogeneity of pyrolyzed carbon-nitride SAC materials~\cite{Intro_x1, Intro_x2}. Understanding how a well-defined MPc coordinates to and behaves on a specific surface is therefore of both fundamental interest and direct relevance to the rational design of single-atom catalysts~\cite{intro_rev_1, Intro_x4}.

Isolating what the central metal ion contributes to any of these properties, however, requires a reference against which the metalated cases can be compared: the metal-free phthalocyanine H$_2$Pc. Its central cavity contains two H atoms bound to a pair of opposite pyrrolic N atoms, reducing the in-plane molecular symmetry from fourfold (D$_\mathrm{4h}$) to twofold (D$_\mathrm{2h}$). The absence of a metal with partially filled d-orbital removes the possibility of a metal-centered frontier-orbital protruding perpendicular to the molecular plane, a pathway through which many MPc couple to their substrates~\cite{MPc_Cu111, CuPcCoPc_STM}, leaving only the protonated macrocycle itself to interact with the surface. Even on metal substrates, H$_2$Pc is less often studied than its various metalated versions. Similar to MPc, however, H$_2$Pc adsorbs mostly flat or slightly bowl-shaped on atomically flat surfaces~\cite{H2PcAg100, H2PcAu111, H2PcCu110-O}, slightly bent on the more corrugated Au(110) surface~\cite{H2PcAu110}, and it is readily metalated by co-deposition of transition-metal atoms or through direct extraction of substrate atoms~\cite{H2PcAg110_metalation, H2PcCu111_metalation, H2PcAg111_metalation}. Moreover, tip- or photo-induced tautomerization has been unambiguously observed between the two pyrrolic NH-orientation configurations~\cite{H2Pc_tauto1, H2Pc_tauto2, H2Pc_tauto3, H2Pc_tauto4}. To our knowledge, analogous studies of H$_2$Pc on well-defined oxide substrates are essentially absent with one exception, \ch{TiO2}(110). In an STM and XPS study of H$_2$Pc on \ch{TiO2}(110)~\cite{H2PcTiO2}, the molecules of a sub-monolayer coverage are found to be preferentially oriented with respect to the surface corrugation, they stay overall intact (no self-metalation) at 300\,K but loose their inner H atoms after annealing to 450\,K.


Indium oxide (\InOx) is a wide-bandgap semiconductor that crystallizes in the cubic bixbyite structure. The (111) facet, its most stable low-index surface~\cite{AgostonAlbe}, hosts 40 atoms per unit cell in a dipole-free, charge-neutral O$_{12}$--In$_{16}$--O$_{12}$ stacking of O$^{2-}$ and In$^{3+}$ ions (Tasker II~\cite{Tasker1979}). The surface structure is a relaxed bulk-termination~\cite{MW2014} with three-fold symmetry and a lattice constant of 1.43\,nm~\cite{InOxlattice}. The surface is composed of two inequivalent types of octahedrally coordinated In(6c) cations and four inequivalent fivefold coordinated In(5c) cations, combined with four inequivalent threefold coordinates O(3c) anions above the In layer, and 12 tetrahedrally coordinates O(4c) sites below. The atomic model of the surface is shown in Fig.\,\ref{fig:InOx_surf} together with an STM image probing the empty states, where the typical pattern of dark triangles related to the low density of states (DOS) of the In(6c) is visible.
\InOx\ is best known as the base of the transparent conductive oxide indium tin oxide (ITO)~\cite{ITO_2, ITO_3}, which makes its organic--oxide interfaces directly relevant to real device architectures~\cite{ITO_organics_interface}. More recently, \InOx\ has emerged as an active catalyst for the CO$_2$-to-methanol hydrogenation reaction, with the (111) facet identified as one of the catalytically most active surfaces~\cite{InOc_cata_Frei2018, InOx_cata_cuyena, InOx_cata_Ding2025}. This provides a direct technological motivation for characterizing how catalytically active molecules including MPcs bind to and organize on this specific surface. The adsorption of cobalt and copper phthalocyanines (CoPc, CuPc) has been recently characterized on the \InOx(111) surface, establishing their site preferences, geometric distortions, ordered structures and electronic properties within the first-layer~\cite{MW2022CoPc, MWCuPc2025}. Crucially, despite their different shapes, the size of the phthalocyanine molecule (1.76 nm$^2$) is essentially the same as the \InOx(111) unit cell (1.77 nm$^2$), therefore a well-ordered (1$\times$1) monolayer (ML) with one molecule per unit cell is geometrically feasible and has been observed for CuPc. In addition to the stoichiometric surface, mildly reducing preparation conditions produce an ordered array of In$^0$ adatoms, with the adatoms located in are B of the unit cell~\cite{MW2014}, see Fig.\ref{fig:InOx_surf}a,b. This reduced surface offers a distinct chemical environment: unlike the substrate cations of the stoichiometric surface, which are locked in bulk lattice positions, the In$^0$ adatoms represent a potentially accessible source of substrate atoms that could, in principle, participate in self-metalation of an adsorbed H$_2$Pc molecule.

The adsorption of the metal-free H$_2$Pc on \InOx(111) is characterized here by low-temperature scanning tunneling microscopy (STM), scanning tunneling spectroscopy (STS), and non-contact atomic force microscopy (nc-AFM), 
as a benchmark against which the effects of the central metal ion can be isolated. Three findings are presented in this work: H$_2$Pc adsorbs exclusively in the previously identified flower (`F') configuration, i.e., in the same site, in-plane orientation and distortions as CuPc and the majority of CoPc~\cite{MW2022CoPc, MWCuPc2025}. Self-metalation of H$_2$Pc is not observed at on the stoichiometric \InOx(111) surface at 300\,K, while on the reduced surface decorated with In$^0$ adatoms different features related to H$_2$Pc were observed. H$_2$Pc starts to decompose on \InOx(111) at $\approx$50\,$^{\circ}$C, precluding the annealing step that would normally be used to prepare well-ordered domains at higher coverages or promote metalation.\\

\section{Experimental methods} 
The experiments were conducted in an Omicron low temperature STM/AFM operated at 4.7\,K. The UHV system used is made up of two chambers, both operating at a base pressure $<$6$\times$10$^{-11}$\,mbar. A qPlus sensor ($k$ = 1800\,N/m, $f_\mathrm{R}$ = 30,010\,Hz, $Q$ = 12,000) in combination with a differential preamplifier and a separate wire for the tunneling current was used. The tip was prepared by electrochemically etching a $\approx$20\,µm thick tungsten wire and gluing it to the sensor's cantilever. In UHV, the tip was initially conditioned on a copper single crystal by applying voltage pulses, as well as dipping it gently into the surface. The final tip preparation was done directly on the H$_2$Pc/\ch{In2O3}(111) surface, where a frequency shift in constant-current STM of less than $-$3\,Hz at a setpoint of +1\,V and 20\,pA and a high lateral resolution in imaging were aimed for. 
The two nominally undoped samples used in this work were: An \ch{In2O3}(111) thin film (crystalline thin film of 200\,nm thickness grown on 5$\times$5$\times$0.5\,mm$^3$ YSZ(111) by pulsed laser deposition~\cite{MWFranceschi2019}), and a single crystal grown with the flux method at Oak Ridge National Laboratory (bulk characterization of a batch with identical surface properties can be found in Ref.~\cite{Hagleitner}). Before depositing H$_2$Pc onto the substrate, the \ch{In2O3}(111) samples were cleaned by repeated cycles of sputtering (Ar$^+$ ions, 1\,keV, 5\,min, rastered across the sample with $\approx$1.6\,µA/cm$^2$) and annealing to $>$300\,$^{\circ}$C in 1$\times$10$^{-6}$\,mbar \ch{O2}. The thin film sample was used for H$_2$Pc adsorption on the stoichiometric (oxidized) surface, the single crystal, which is more robust upon thermal reduction was used for adsorption on the reduced surface. Thus, for the stoichiometric \ch{In2O3}(111) surface, the thin film sample was left in an oxygen-rich atmosphere until it cooled to $\approx$150\,$^{\circ}$C to prevent the surface from reducing. To prepare the reduced \ch{In2O3}(111) surface with ordered In adatoms, the final annealing of the single crystal took place in UHV. A water-cooled 4-pocket evaporator from Omnivac was used to thermally sublimate the H$_2$Pc powder (Sigma-Aldrich, $\beta$-form, 98\% dye content). In preparation for molecule evaporation, the H$_2$Pc powder was thoroughly degassed by annealing for more than 12 hours in the UHV system. This procedure was done to reduce the intrinsically present water in the powder, which would otherwise co-adsorb on the \ch{In2O3}(111)~\cite{MW2017water}. Scanning tunneling microscopy was performed in constant current mode, atomic force microscopy in constant height, non-contact mode. In the AFM images attractive forces are characterized by a negative frequency shift (dark) and repulsive forces by a more positive frequency shift (bright) with respect to the attractive background. Image processing was done with the free software ImageJ. H$_2$Pc coverages were determined by counting the molecules in several 20 and 40\,nm-sized images and averaging the numbers. Differential conductance spectra (d$I$/d$V$) on the H$_2$Pc molecules and the \ch{In2O3}(111) surface were acquired using a lock-in amplifier (Zurich Instruments) at 137\,Hz with an amplitude of 10\,mV. The setpoint for spectroscopy (+1.0\,V, 15\,pA) positioned the tip rather far from the surface; measuring at smaller tip--sample distances led to irreversible changes of the H$_2$Pc molecules. Several spectra were averaged and final smoothing was applied using Savitzky–Golay filtering.

\section{Results}

\subsection{Isolated H$_2$Pc molecules}

\begin{figure}[H]
    \centering
    \includegraphics[width=\linewidth]{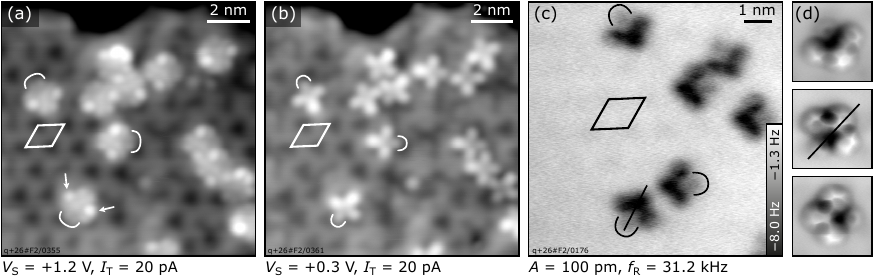} 
    \caption{Individual H$_2$Pc molecules on \InOx(111). (a, b) STM images of the same surface area acquired at the LUMO energy of +1.0\,V and in the H$_2$Pc HOMO--LUMO gap at +0.3\,V, respectively. The surface unit cell is drawn with solid white (black) lines with corners located at the 3-fold rotation axis B. White (black) curves mark the isoindole group of the H$_2$Pc bent towards the surface for the three rotational orientations. Arrows indicate the two brightest lobes of a molecule as a guide to identify the rotational orientations of the molecule. (c) AFM image demonstrating the different brightness/contrast across the H$_2$Pc molecules. Note that this image was taken on a different surface area than panels a and b. (d) AFM images of three molecules at small tip--molecule distances revealing the internal structure of the molecules.}
    \label{fig:single_H2PC}
\end{figure}

Fig.\,\ref{fig:single_H2PC}a,b shows STM images of the stoichiometric \ch{In2O3}(111) surface after depositing $\approx$0.2\,ML of H$_2$Pc at 300\,K (1\,ML is defined here as one molecule per substrate unit cell). Individual molecules are randomly distributed on the surface with no preference for step edges or defects (see Supplement), and the pattern of dark features characteristic for the bare \ch{In2O3}(111) surface is seen in-between the molecules. In contrast to many metal surfaces, Pc molecules do not diffuse at 300\,K on \ch{In2O3}(111) and can be imaged as static objects at room temperature (see Supplement). When tunneling into the lowest unoccupied molecular orbital (LUMO) of the molecule (Fig.\,\ref{fig:single_H2PC}a), which is located at $\approx$+1.0\,eV, each H$_2$Pc molecule is imaged as a flower-like object with eight bright lobes arranged in pairs around the center. Two specific lobes across the molecule are brighter than all others as indicated by arrows in Fig.\,\ref{fig:single_H2PC}a. The three orientations rotated by 120$^\circ$ relative to each other (reflecting the surface symmetry) can be discerned even though the orbital structure obscures most of the internal structure of the molecules. It should be noted that the adsorption of H$_2$Pc with two-fold symmetry (D$_{2\mathrm{h}}$) in the gas phase onto a surface with three-fold symmetry (p3) results in a combined system where the relaxed molecule has lost all symmetry. This lack of symmetry is visible in the internal structure of H$_2$Pc when tunneling into the LUMO (see high-pass filtered STM images in the Supplement). The three orientations are identified more clearly in the STM images acquired at +0.3\,V shown in Fig.\,\ref{fig:single_H2PC}b, i.e., by tunneling in the HOMO--LUMO gap of the molecule. Here, the H$_2$Pc resembles a cross-shaped feature with lobes (isoindole units) of different apparent lengths. To investigate the adsorption configuration of H$_2$Pc further, constant-height nc-AFM is employed as it is more sensitive to the internal structure of organic molecules than STM.

Fig.\,\ref{fig:single_H2PC}c presents an AFM image acquired at a tip-sample distance where only long-range and short-range attractive forces contribute, revealing different brightness (frequency shift) on the four isoindole units of the molecule: three appear darker (indicating a stronger attractive interaction with the tip) than the fourth, suggesting that the three dark ones protrude further from the surface than the remaining one. Comparison of the STM and AFM images identifies the isoindole group closest to the surface as the short lobe seen in the cross-shaped STM feature at +0.3\,V (Fig.~\ref{fig:single_H2PC}b). Bringing the AFM tip closer to the surface and into the regime where short-range repulsive forces dominate, shown in Fig.\,\ref{fig:single_H2PC}d, resolves the intramolecular chemical structure of H$_2$Pc. The three protruding isoindole units appear more repulsive (brighter) and show sharp features of the phenyl rings (comparable to imagining with a flexible tip termination like a CO molecule) compared to the fourth, which is less visible and separated from the rest of the molecule by an asymmetric dark region spreading from the center of the molecule. This strong contrast variation across the center (see black line across one of the molecule in panel d) prevents an unambiguous assignment of the inner H positions: whether the two H atoms sit on the pair of pyrrolic N atoms across the distortion (along the black line) or on the pair parallel to it cannot be determined from the AFM data.

STM images where the underlying \InOx(111) lattice is visible between the molecules as it is the case in those shown in Fig.\,\ref{fig:single_H2PC}a,b (despite the presence of a few OH groups), are suitable to triangulate the adsorption site and orientation of the molecules (see example in the Supplement). Fig.\,\ref{fig:exp_site} shows the result: The center of the molecule (where the metal atom of an MPc would be) is located close to an O($\gamma$) anion of the substrate, possibly slightly displaced toward the neighboring In(c) cation. The rotational orientation of the molecule is such that the In(6c) region (around the B axis, blue In atoms in Fig.~\ref{fig:exp_site}) are largely avoided; only one iminic N atom is positioned close to the O($\beta$) anion of the In(6c) region nearest to the O($\gamma$) of the molecule's center. Consequently, two opposite isoindole units are approximately aligned along a $\langle 1\overline{1}0 \rangle$ direction, and these are the lobes located across the observed molecular distortion (along the black line across the molecule in Fig.\,\ref{fig:single_H2PC}c,d). This adsorption geometry and site corresponds to the flower (`F') configuration previously identified for cobalt and copper phthalocyanines on this surface~\cite{MW2022CoPc, MWCuPc2025}.

\begin{figure}[H]
    \centering
    \includegraphics[width=0.5\linewidth]{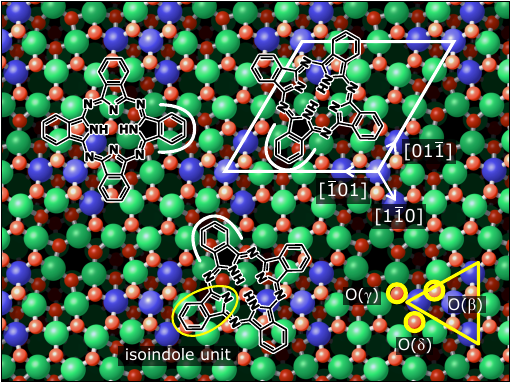}
    \caption{Schematic of the H$_2$Pc adsorption site on \InOx(111) extracted from the STM and AFM data. This figure uses the undistorted gas-phase structure of the molecule superimposed onto the relaxed structure of the bare \InOx(111) surface.}
    \label{fig:exp_site}
\end{figure}

To locate the frontier orbitals of H$_2$Pc relative to the electronic structure of the substrate, scanning tunneling spectroscopy (STS) measurements were performed on both, the molecules and the bare \InOx(111) surface, as shown in Fig.\,\ref{fig:STS}a. As introduced above, \ch{In2O3} has a fundamental band gap of $\approx$2.9\,eV (see Supplement) and the material is intrinsically n-doped, with the Fermi level (E$_\mathrm{F}$) located at the conduction band minimum (dashed line in Fig.\,\ref{fig:STS}a). Against this background, the HOMO of the H$_2$Pc appears at $\approx-$1.7\,eV, well inside the band gap of the substrate, and the LUMO appears at $\approx$+1.0\,eV, i.e., $\approx$1.0 eV above the conduction band minimum and therefore energetically overlapping with the substrate conduction band states. Both features are somewhat broadened in the STS spectra; whether this broadening reflects a genuine hybridization of the LUMO with the In\,5s-derived conduction band states~\cite{InOx_DFT_bandstructure} or has an instrumental origin is not resolved by the present measurement. In either case, the LUMO sits at $\approx$1 eV above the Fermi level, i.e., far enough from E$_\mathrm{F}$ that charge transfer between the molecule and the substrate is not expected, and consistently, the STS spectra show no features around E$_\mathrm{F}$ that would indicate partial LUMO occupation. Constant-current STM images acquired at bias voltages corresponding to the HOMO and LUMO shown in Fig.\,\ref{fig:STS}b complement the energetic picture by mapping the spatial distribution of the frontier orbitals. Both orbitals appear as eight-lobed ``flower'' patterns in which each isoindole unit of the H$_2$Pc molecule contributes a pair of lobes, so that all four isoindoles are electronically visible. As mentioned before, two opposite lobes are systematically brighter than the other lobes both in the HOMO and LUMO, and the isoindole unit pointing towards the surface appears slightly darker than the opposite one. The nodal planes and the modulation of the lobe brightness produces an apparent two-fold pattern with a mirror plane along the $\langle 1\overline{1}0 \rangle$ direction, i.e., the axis passing through the two isoindoles across the vertical distortion. In the LUMO image, the same modulation is present, though the lobe pairs are less sharply resolved than in the HOMO. A closer look reveals that this apparent two-fold symmetry is not exact. A residual asymmetry is visible at the center of both HOMO and LUMO images (see Supplement for high-pass filtered STM images), breaking the mirror plane that would formally preserve D$_{2\mathrm{h}}$. This is consistent with the H$_2$Pc molecule sitting on a surface that lacks mirror symmetry, and the approximately two-fold appearance in STM and AFM reflects the residual molecular symmetry rather than a symmetry of the adsorbed system as a whole.

\begin{figure}[H]
    \centering
    \includegraphics[width=0.5\linewidth]{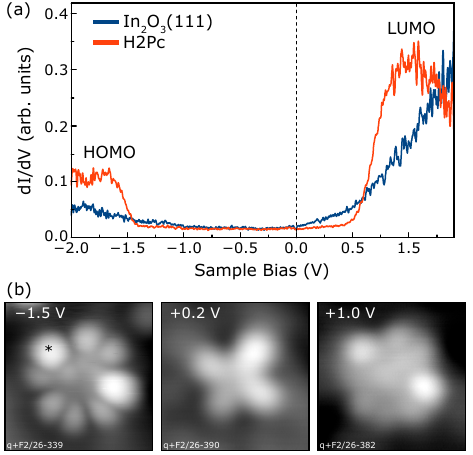}
    \caption{Local electronic structure of H$_2$Pc on \ch{In2O3}(111). (a) Differential conductance (d$I$/d$V$) spectra taken on the bright lobe of the H$_2$Pc molecules (red trace) and the bare \ch{In2O3}(111) surface (blue trace) next to the molecule. (b) STM images acquired at the energy of the HOMO, LUMO, and in the HOMO--LUMO gap. The position on the molecule were the d$I$/d$V$ spectra were taken is indicated by an asterisk.}
    \label{fig:STS}
\end{figure}

\subsection{Increasing coverage and thermal stability of H$_2$Pc}

\begin{figure}[H]
    \centering
    \includegraphics[width=\linewidth]{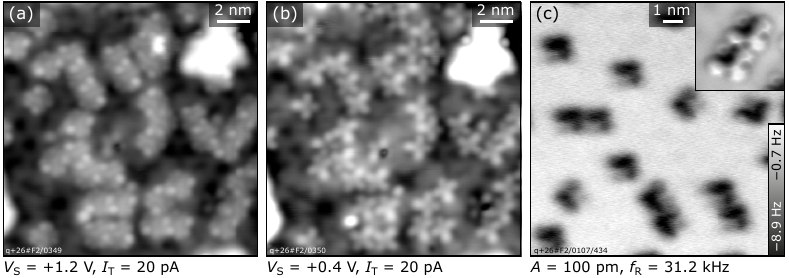} 
    \caption{Pairs formed by 0.2\,ML of H$_2$Pc molecules on \InOx(111). (a, b) STM images acquired at the LUMO energy (+1.0\,V) and in the H$_2$Pc HOMO--LUMO gap (+0.4\,V). (c) AFM image showing three pairs. This image was taken on a different surface area than (a, b). (d) AFM images of a H$_2$Pc pair at small tip--molecule distances showing the internal structure of the molecules.}
    \label{fig:chains}
\end{figure}

At coverages of $\approx$0.2--0.5\,ML per unit cell, H$_2$Pc populates the surface in the F configuration described above, by arranging into pairs or short chains along $\langle 1\overline{1}0 \rangle$ directions, see Fig.\,\ref{fig:chains}. The individual molecules within these chains adopt one of the three symmetry-equivalent F orientations, with all molecules of a chain adapting the same orientation. The site and thus also the separation of the molecules within a chain is enforced by the \ch{In2O3}(111) lattice and no steric hindrance or overlap occurs between molecules. Each molecular orientation can form chains in two $\langle 1\overline{1}0 \rangle$ directions, where the surface atoms between the molecules are either In(6c) or In(5c), as discussed in Ref.~\cite{MWCuPc2025}. STM images acquired at the LUMO energy, depicted in Fig.\,\ref{fig:chains}a, and at +0.4\,V, see Fig.~\ref{fig:chains}b, show the same intramolecular features as for the isolated molecules (cross-shaped feature with a short arm identifying the non-protruding isoindole), and AFM images at intermediate and close tip-sample distances, Fig.~\ref{fig:chains}c, confirm that the molecular distortion and adsorption geometry are preserved in this arrangement. 

\begin{figure}[H]
    \centering
    \includegraphics[width=0.65\linewidth]{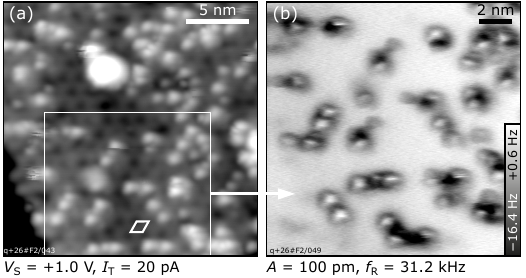} 
    \caption{Thermal stability of H$_2$Pc on \ch{In2O3}(111). (a) STM and (b) AFM images of a submonolayer coverage of H$_2$Pc after annealing to 250\,$^\circ$C, revealing molecular fragments.}
    \label{fig:stability}
\end{figure}

The pair- and chain-formation regime closely parallels what has been reported for CoPc and CuPc on \ch{In2O3}(111)~\cite{MW2022CoPc, MWCuPc2025}. For these MPcs, further deposition combined with a moderate thermal treatment drives the coverage-dependent progression from chains through a (2$\times$2) intermediate structure at 0.75\,ML to a fully ordered (1$\times$1) monolayer with one molecule per substrate unit cell. Attempts to reproduce this preparation for H$_2$Pc, however, failed: heating a partial H$_2$Pc layer to 250\,$^\circ$C (the temperature routinely used for CoPc and CuPc without any indication of chemical change or degradation of the molecules) leads to complete fragmentation of all H$_2$Pc molecules, visible as small, irregular features on the surface instead of the intact cross-shaped molecule, see Fig.\,\ref{fig:stability}. In AFM, a variety of small objects is observed but isoindole units or phenyl-ring related features could not be identified. In the attempt to provide less thermal energy, it was found that heating to only $\approx$50\,$^\circ$C already produces fractured molecules (see Supplement). It can be concluded that H$_2$Pc is structurally unstable on \ch{In2O3}(111) at temperatures well below the reported sublimation temperature of $\approx$380\,$^\circ$C of the pure powder, and that annealing cannot be used to promote diffusion and the formation of larger ordered domains.

\begin{figure}[H]
    \centering
    \includegraphics[width=0.65\linewidth]{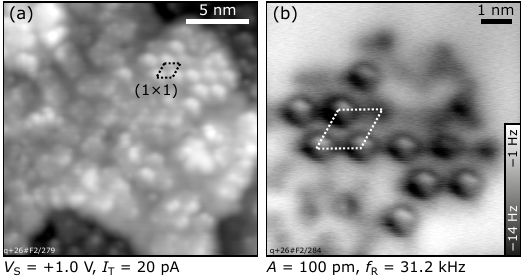} 
    \caption{Increasing the H$_2$Pc coverage. (a) Deposition of $>$0.6\,ML of H$_2$Pc on \ch{In2O3}(111) result in a surface with seemingly disordered molecules and only small patches of an ordered (1$\times$1) structure. (b) AFM image of the (1$\times$1) domain.}
    \label{fig:1x1}
\end{figure}

Consequently, higher coverages of H$_2$Pc on \ch{In2O3}(111) can only be prepared by deposition at 300\,K, where diffusion does not take place (see RT-STM images in the Supplement). At a coverage of $>$0.6\,ML, small (1$\times$1) domains of no more than about ten molecules each are found, while the majority of the surface is covered by intact isolated H$_2$Pc forming short chains and populating the surface in a disordered way, as shown in Fig.\,\ref{fig:1x1}a. The molecules within a (1$\times$1) domain are all oriented the same way and only the protruding isoindoles are visible in STM as bright double feature, matching with what has been previously observed for MPc~\cite{MW2022CoPc, MWCuPc2025}. Fig.\,\ref{fig:1x1}b provides an AFM images of the small (1$\times$1) patches, where only the protruding isoindoles are visible. The (2$\times$2) intermediate structure observed for CuPc and CoPc at 0.75\,ML is not observed: this phase consists of a building block made of three molecules in different orientations, i.e., it requires the molecules to substantially rearrange, and the diffusion barrier for such rearrangement cannot be overcome at 300\,K.

\subsection{Self-metalation on \ch{In2O3}(111)?}

On the stoichiometric \ch{In2O3}(111) surface at 300\,K, no evidence of self-metalation is found. In the STM data, H$_2$Pc and CuPc~\cite{MWCuPc2025} are almost indistinguishable in appearance (compared to CoPc~\cite{MW2022CoPc}), and the distortion of the molecules makes it difficult to access the structure in the cavity with AFM. The decomposition of H$_2$Pc on \ch{In2O3}(111) at temperatures where MPcs are not affected, strongly indicates that the molecules of the present study are indeed H$_2$Pc.

\begin{figure}[H]
    \centering
    \includegraphics[width=\linewidth]{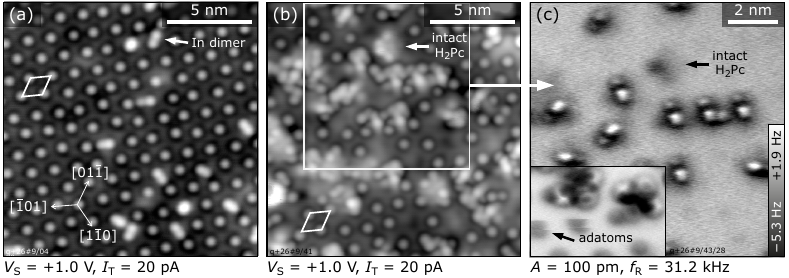}  
    \caption{H$_2$Pc on reduced \InOx(111). (a) STM image of the reduced surface, where In adatoms located at the symmetry point B of the unit cell form an ordered array. A few In dimers are present as well. (b) H$_2$Pc on the reduced surface, imaged with STM. Individual molecules resembling those imaged on the bare surface are found, but the majority have a changed appearance. (c) AFM image of an area of panel (b) revealing that the features are still well-defined and present in three orientations.}
    \label{fig:reduced}
\end{figure}

To evaluate whether metal atoms available on the surface would change this picture, the reduced \ch{In2O3}(111) surface bearing an ordered array of In$^0$ adatoms~\cite{MW2014} was prepared. The In adatoms are immobile at 300\,K in the sense that they do not diffuse across the surface even at a partial coverage. They do, however, show local mobility between the three bridge positions around the B axis. The full monolayer with one In adatom per unit cell was prepared as shown in Fig.~\ref{fig:reduced}a. Deposition of H$_2$Pc onto the reduced surface yields a mixed adsorbate configuration as shown in Fig.~\ref{fig:reduced}b,c. Only a few intact H$_2$Pc molecules in the F configuration described above are found, while the majority of features no longer resemble a phthalocyanine either in STM or in AFM. These features still appear in three symmetry-equivalent rotational orientations on the surface, suggesting they are well-defined objects in a specific site rather than random decomposition products, but their internal structure is inaccessible with both STM and AFM. Compared to regularly-adsorbed H$_2$Pc or the In adatoms on the surface (see inset in Fig.~\ref{fig:reduced}c), these features protrude much further from the surface, suggesting that they may be strongly tilted molecules with In adatoms nearby or underneath.

\section{Discussion}

The observation that H$_2$Pc occupies the same F site, in-plane orientation, and out-of-plane distortions as CoPc and CuPc do in F~\cite{MW2022CoPc, MWCuPc2025} establishes the F configuration as the intrinsic preference of the phthalocyanine macrocycle on \ch{In2O3}(111). Since H$_2$Pc contains no transition-metal ion, the F site cannot be selected by any  metal–substrate interaction and must arise from the interaction of the organic macrocycle itself with the surface. Correspondingly, the square S configuration observed exclusively for CoPc~\cite{MW2022CoPc} is confirmed as a genuine metal-specific effect. Whether in F site the two inner H atoms sit on the pair of pyrrolic N atoms across or parallel to the vertical distortion cannot be determined from the present STM and AFM data. Throughout the whole coverage regime, the metal-free molecule behaves as its metalated analogues in F site (with the restriction to 300\,K and thus limited diffusion that prevents (2$\times$2) arrangement), which is to be expected since the adsorption site remains the same. 

The absence of the central metal appears directly connected to the fragmentation of H$_2$Pc at slightly elevated temperatures. In the metalated MPcs, the four isoindole units are held together by two independent bonds: the iminic N atoms of the macrocycle around the ring, and the direct coordination of all four inner (pyrrolic) N atoms to the central M$^{2+}$ cation. In H$_2$Pc only the iminic N bonds remain, and on \ch{In2O3}(111) these prove insufficiently strong to prevent disintegration at moderately elevated temperatures, i.e., well below the $\approx$380\,$^\circ$C sublimation temperature of pure H$_2$Pc. On rutile \ch{TiO2}(110), H$_2$Pc adsorbs flat and centered on the substrate O rows, and while heating leads to dehydrogenation of the two H atoms in the central cavity, the macrocycle framework itself is preserved up to at least 450\,K~\cite{H2PcTiO2}. On \ch{In2O3}(111), by contrast, fragmentation starts already at 50\,$^\circ$C, a much lower temperature and a more drastic outcome. This suggests that the \ch{In2O3}(111) surface plays an active role in the decomposition, though the specific chemistry involved remains open.


Whether the substrate provides atoms that can be captured into the H$_2$Pc cavity to form a metalated MPc was one of the questions addressed in this work. In general, two routes are possible: (1) extraction of a lattice atom, or (2) incorporation of an adatom. On many metal substrates, self-metalation is driven by the mobility of substrate atoms and the low barrier for their capture into the phthalocyanine cavity: on Cu(111) metalation starts already at 240\,K~\cite{H2PcCu111_metalation}, on Ag(110) at 300\,K~\cite{H2PcAg110_metalation}. H$_2$Pc remains pristine on Ag(111) even after annealing at 530\,K, but metalation with post-deposited Fe was observed at 300\,K~\cite{H2PcAg111_metalation}. On an oxide surface such as \ch{In2O3}(111) or \ch{TiO2}(110) the mechanistic picture is different in general: the substrate cations are held more rigidly in their lattice positions by directional ionic-covalent bonds and are not readily available from step edges or defects for incorporation into the molecular cavity. Self-metalation on an oxide surface is therefore not per se expected.

Beyond this general argument, In specifically is a poor candidate for the H$_2$Pc cavity. The cation of a standard MPc sits in a planar four-fold coordination and in the M$^{2+}$ oxidation state, whereas In (like early transition metals) prefers higher coordination and the In$^{3+}$ oxidation state. An In cation in an H$_2$Pc cavity would therefore require an additional ligand to satisfy its coordination preferences, which are typically halogens~\cite{InPc_halogens}. Consistently, the commercially available halogenated specimen is Indium(III) phthalocyanine chloride, which loses its Cl ligand when evaporated onto Pb(100), and those InPc molecules in which the In cation points above the molecular plane exhibits Yu-Shiba-Rusinov states signaling a localized magnetic spin~\cite{InPc_1}. On \ch{In2O3}(111) at 300\,K, the first route (extraction of an In cation) seems to be not favorable: incorporation of a lattice In from the stoichiometric surface would require a coordination change without adequate compensation (In cannot form a (M=O)$^{2+}$ compound equivalent to titanyl or vanadyl by extracting a lattice O) and cleaving off the H atoms from the molecule (which could protonate surface oxygen atoms), energetically too costly to be beneficial. The second route via the metallic In$^{0}$ adatoms produced upon mild surface reduction, while chemically accessible, remains unclear. The reduced surface yields a mixed adsorbate of intact F-configured H$_2$Pc (minority) together with distorted features that do not resemble a Pc molecule any more and thus cannot be identified by imaging alone.

Due to the lack of experimental literature for H$_2$Pc on oxide surfaces apart from Ref.~\cite{H2PcTiO2}, the behavior of H$_2$Pc is compared to that of the structurally closely related free-base tetraphenylporphyrin (2HTPP) on oxide surfaces. On rutile \ch{TiO2}(110)~\cite{2HTPP_ox_TiO2_1}, the molecules of the first layer adsorb at room temperature oriented but without long-range ordering and they are protonated by surface protons forming the diacid 4HTPP$^{2+}$. (Whether protonation occurs depends strongly on the availability of surface protons or H atoms dissolved in the bulk and was not observed in Ref.~\cite{2HTPP_ox_TiO2_1a}.) 2HTPP molecules of the second layer self-metalate upon heating to $\approx$270\,K form titanyl tetraphenylporphyrin (TiO-TPP), with the reaction becoming complete and all molecules metalated at ~550\,K. When a substrate cation is incorporated into the molecule, Ti$^{2+}$ is an unusual oxidation state and the most common ligand to reach Ti$^{3+}$ is oxygen to form (Ti=O)$^{2+}$. The proposed reaction mechanism for metalation considers an Ti interstitial that is extracted from the surface, and in the ion exchange the two protons from the molecule compensate the loss of (Ti=O)$^{2+}$ from the surface~\cite{2HTPP_ox_TiO2_2}. Since 2HTPP does not self-metalate on \ch{TiO2}(110) at room temperature, metalation by post-deposited metals is possible, for example with Ni~\cite{2HTPP_ox_TiO2_Ni_1, 2HTPP_ox_TiO2_Ni_2}. Self-metalation reactions have also been reported on MgO(001) thin films, where the reaction takes place at room temperature promoted by step edges and defects~\cite{2HTPP_ox_MgO_1}. Similarly, self-metalation was observed on Co-terminated Co$_3$O$_4$(111) and oxygen-terminated CoO(111) starting at 275\,K and still being observed at 475\,K~\cite{2HTPP_ox_CoO_1, 2HTPP_ox_CoO_2}.

A consistency is observed for H$_2$Pc and 2HTPP on \ch{TiO2}(110), which both do not self-metalate at room temperature. The systems differ in the fact that H$_2$Pc deprotonates upon heating to 450\,K but apparently does not metalate, while 2HTPP gets protonated at room temperature and self-metalates at elevated temperatures. The origin of this difference has not been systematically explored. One can only speculate that the different reactivity of H$_2$Pc and 2HTPP for de-/protonation is related to the spacial localization of the HOMO and LUMO around the cavity which differs (in orbital calculations) due to the structural differences in the pyrroles (fused with benzene in the case of H$_2$Pc, bare pyrrols for 2HTPP) and the connection of the four pyrrols (via iminic --N= in the case of H$_2$Pc, and methine bridges --CH= with phenyl rings for 2HTPP).

Three features consistently distinguish these systems from ours: (i) the molecules are stable on the oxide surfaces at elevated temperatures; (ii) 2HTPP either metalates directly at room temperature or at slightly elevated temperatures depending on the oxide surface; (iii) the substrate metal (Ti, Mg, Co) forms a chemically favorable 4-coordinate M--N$_4$ complex or an (M=O)$^{2+}$ to increase the otherwise low oxidation state.

\section{Summary}
H$_2$Pc and MPc are often studied on metal surfaces, but systematic studies on oxide surfaces are missing with the exception of \ch{TiO2}(110)~\cite{H2PcTiO2}. Moreover, the role of the metal ion in the interaction with a surface is rarely separated from the interaction of the metal-free but protonated macrocycle. In this work the adsorption of metal-free H$_2$Pc was characterized on the (111) surface of \ch{In2O3} using STM, nc-AFM, and STS at low temperature. H$_2$Pc on \ch{In2O3} provides a benchmark case allowing us to compare its adsorption with previously studied metal phthalocyanines, CuPc and CoPc. Individual H$_2$Pc molecules are found to occupy the same adsorption site and configuration (flower `F'), as CuPc and (to some extent) also CoPc. The population of the the F configuration regardless of the presence of a transition-metal ion reveals the preference of the phthalocyanine ligand itself, driven by the interaction of the aromatic macrocycle with the substrate rather than by any coupling between a central metal and the surface below it.

The consequences of the missing metal appear once the coverage is increased or the surface is gently heated. Neither CoPc nor CuPc show any chemical change on annealing at 250\,$^\circ$C, the standard treatment to promote diffusion and grow their extended and well-defined phases of the first layer. H$_2$Pc, by contrast, starts to disintegrate already at $\approx$50\,$^\circ$C on \ch{In2O3}(111), which limits the structures and their long-range ordering that can be formed at higher coverages. Moreover, self-metalation is not observed on the stoichiometric \ch{In2O3}(111) surface, while on the reduced surface (decorated with metallic In adatoms) new specimen are observed but their nature remains unclear.

Taking these observations together, the central metal ion emerges as a component of the MPc/\ch{In2O3}(111) system with three separable roles: it influences the character of the frontier orbitals (ligand-centered versus metal--ion-centered LUMO); it stabilizes the macrocycle against decomposition on this surface; and, in the specific case of CoPc, it opens a second adsorption site by coupling directly to the In cation below due to hybridization of the d$_{z^2}$-type LUMO with conduction band of the surface. This work on H$_2$Pc establishes that the central metal ion is a prerequisite for stability against fragmentation at mildly elevated temperatures, and thus the preparation of extended ordered MPc films on \ch{In2O3}(111).

\setstretch{1.00}

\section*{Acknowledgement}
This research was funded in part by the Austrian Science Fund (FWF) the Cluster of Excellence MECS [10.55776/COE5], and [10.55776/V773]. For open access purposes, the authors have applied a CC BY public copyright license to any author accepted manuscript version arising from this submission.

\bibliographystyle{unsrt}
\bibliography{references}

@article{InOx_DFT_bandstructure,
  title = {{Electronic band structure of indium tin oxide and criteria for transparent conducting behavior}},
  author = {Mryasov, O. N. and Freeman, A. J.},
  journal = {Phys. Rev. B},
  volume = {64},
  issue = {23},
  pages = {233111},
  numpages = {3},
  year = {2001},
  month = {Dec},
  publisher = {American Physical Society},
  doi = {10.1103/PhysRevB.64.233111},
  url = {https://link.aps.org/doi/10.1103/PhysRevB.64.233111}
}

@article{InOc_cata_Frei2018,
   author = {M. S. Frei and M. Capdevila-Cortada and R. García-Muelas and C. Mondelli and N. López and J. A. Stewart and D. Curulla Ferré and J. Pérez-Ramírez},
   doi = {10.1016/j.jcat.2018.03.014},
   issn = {10902694},
   journal = {Journal of Catalysis},
   month = {5},
   pages = {313-321},
   publisher = {Academic Press Inc.},
   title = {{Mechanism and microkinetics of methanol synthesis via CO$_2$ hydrogenation on indium oxide}},
   volume = {361},
   year = {2018}
}

@article{Intro_x4,
    author = {Dehui Deng  and Xiaoqi Chen  and Liang Yu  and Xing Wu  and Qingfei Liu  and Yun Liu  and Huaixin Yang  and Huanfang Tian  and Yongfeng Hu  and Peipei Du  and Rui Si  and Junhu Wang  and Xiaoju Cui  and Haobo Li  and Jianping Xiao  and Tao Xu  and Jiao Deng  and Fan Yang  and Paul N. Duchesne  and Peng Zhang  and Jigang Zhou  and Litao Sun  and Jianqi Li  and Xiulian Pan  and Xinhe Bao },
    title = {{A single iron site confined in a graphene matrix for the catalytic oxidation of benzene at room temperature}},
    journal = {Science Advances},
    volume = {1},
    number = {11},
    pages = {e1500462},
    year = {2015},
    doi = {10.1126/sciadv.1500462},
}

@article{intro_rev_1,
    title = {{Advances in coordination engineering of M--N--C single atom catalysts for superior oxygen reduction performance}},
    journal = {Coordination Chemistry Reviews},
    volume = {549},
    pages = {217244},
    year = {2026},
    issn = {0010-8545},
    doi = {https://doi.org/10.1016/j.ccr.2025.217244},
    url = {https://www.sciencedirect.com/science/article/pii/S0010854525008148},
    author = {Anuj Kumar and Naina Goyal and Sanjay Mathur and Ibragimov Aziz Bakhtiyarovich and Yufeng Zhao and Mohammad Khalid and Mohd Ubaidullah and Abdullah M. Al-Enizi},
}

@article{Intro_x2,
    author = {Liu, Wengang and Zhang, Leilei and Liu, Xin and Liu, Xiaoyan and Yang, Xiaofeng and Miao, Shu and Wang, Wentao and Wang, Aiqin and Zhang, Tao},
    title = {{Discriminating Catalytically Active FeN$_x$ Species of Atomically Dispersed Fe--N--C Catalyst for Selective Oxidation of the C--H Bond}},
    journal = {Journal of the American Chemical Society},
    volume = {139},
    number = {31},
    pages = {10790-10798},
    year = {2017},
    doi = {10.1021/jacs.7b05130},
}

@article{Intro_x1,
    title = {Advances of carbon nitride based atomically dispersed catalysts from single-atom to dual-atom in advanced oxidation process applications},
    journal = {Coordination Chemistry Reviews},
    volume = {505},
    pages = {215693},
    year = {2024},
    issn = {0010-8545},
    doi = {https://doi.org/10.1016/j.ccr.2024.215693},
    url = {https://www.sciencedirect.com/science/article/pii/S0010854524000390},
    author = {Jie Deng and Yuxi Zeng and Eydhah Almatrafi and Yuntao Liang and Zihao Wang and Ziwei Wang and Biao Song and Yanan Shang and Wenjun Wang and Chengyun Zhou and Guangming Zeng},
}

@article{InOx_cata_Ding2025,
   author = {Yishui Ding and Jie Chen and Haihong Zheng and Yalong Jiang and Linbo Li and Xiangrui Geng and Xu Lian and Lu Yang and Ziqi Zhang and Kelvin Hongliang Zhang and Hexing Li and Jian Qiang Zhong and Wei Chen},
   doi = {10.1021/prechem.5c00005},
   issn = {27719316},
   journal = {Precision Chemistry},
   publisher = {American Chemical Society},
   title = {{Enhanced Hydrogen Adsorption on In$_2$O$_3$(111) via Oxygen Vacancy Engineering}},
   year = {2025}
}

@article{InOx_cata_cuyena,
    author = {Mathiesen, Jette K. and Zhu, Jie and Wan, Weiming and Shaikhutdinov, Shamil and Roldan Cuenya, Beatriz},
    title = {{Metal–Support Interaction in In$_2$O$_3$‑Based Catalysts of CO$_2$ Hydrogenation Studied Using ``Inverse''' In$_2$O$_3$(111)/Ru(0001) Model Systems}},
    journal = {The Journal of Physical Chemistry C},
    volume = {129},
    number = {18},
    pages = {8582-8590},
    year = {2025},
    doi = {10.1021/acs.jpcc.5c00272},
}

@article{InPc_halogens,
    author = {Sun, Wenfang and Wang, Gang and Li, Yunjing and Calvete, Mario J. F. and Dini, Danilo and Hanack, Michael},
    title = {{Axial Halogen Ligand Effect on Photophysics and Optical Power Limiting of Some Indium Naphthalocyanines}},
    journal = {The Journal of Physical Chemistry A},
    volume = {111},
    number = {17},
    pages = {3263--3270},
    year = {2007},
    doi = {10.1021/jp071152k},
}

@article{2HTPP_ox_CoO_2,
    author = {Wang, Can and Wang, Ruimei and Hauns, Jakob and Fauster, Thomas},
    title = {{Self-Metalation of Porphyrins by Cobalt Oxide: Photoemission
Spectroscopic Investigation}},
    journal = {The Journal of Physical Chemistry C},
    volume = {124},
    number = {26},
    pages = {14167--14175},
    year = {2020},
    doi = {10.1021/acs.jpcc.0c01722},
}

@article{2HTPP_ox_CoO_1,
    author = {Wechsler, Daniel and Fernández, Cynthia C. and Tariq, Quratulain and Tsud, Nataliya and Prince, Kevin C. and Williams, Federico J. and Steinrück, Hans-Peter and Lytken, Ole},
    title = {{Interfacial Reactions of Tetraphenylporphyrin with Cobalt-Oxide Thin Films}},
    journal = {Chemistry – A European Journal},
    volume = {25},
    number = {57},
    pages = {13197--13201},
    doi = {https://doi.org/10.1002/chem.201902680},
    year = {2019}
}

@article{2HTPP_ox_MgO_1,
    author = {Egger, Larissa and Hollerer, Michael and Kern, Christian S. and Herrmann, Hannes and Hurdax, Philipp and Haags, Anja and Yang, Xiaosheng and Gottwald, Alexander and Richter, Mathias and Soubatch, Serguei and Tautz, F. Stefan and Koller, Georg and Puschnig, Peter and Ramsey, Michael G. and Sterrer, Martin},
    title = {{Charge-Promoted Self-Metalation of Porphyrins on an Oxide Surface}},
    journal = {Angewandte Chemie International Edition},
    volume = {60},
    number = {10},
    pages = {5078--5082},
    doi = {https://doi.org/10.1002/anie.202015187},
    year = {2021}
}

@article{2HTPP_ox_TiO2_Ni_2,
    author = {Wang, Cici and Fan, Qitang and Han, Yong and Martínez, José I. and Martín-Gago, José A. and Wang, Weijia and Ju, Huanxin and Gottfried, J. Michael and Zhu, Junfa},
    title = {{Metalation of tetraphenylporphyrin with nickel on a TiO$_2$(110)-1 $\times$ 2 surface}},
    journal = {Nanoscale},
    volume = {8},
    number = {2},
    pages = {1123-1132},
    year = {2016},
    doi = {10.1039/c5nr03134f},
}

@article{2HTPP_ox_TiO2_Ni_1,
    author = {Wang, Cici and Fan, Qitang and Hu, Shanwei and Ju, Huanxin and Feng, Xuefei and Han, Yong and Pan, Haibin and Zhu, Junfa and Gottfried, J. Michael},
    title = {{Coordination reaction between tetraphenylporphyrin and nickel on a TiO$_2$(110) surface}},
    journal = {Chemical Communications},
    volume = {50},
    number = {61},
    pages = {8291-8294},
    year = {2014},
    month = {08},
    doi = {10.1039/c4cc02919d},
}

@article{2HTPP_ox_TiO2_1a,
    author = {Lovat, Giacomo and Forrer, Daniel and Abadia, Mikel and Dominguez, Marcos and Casarin, Maurizio and Rogero, Celia and Vittadini, Andrea and Floreano, Luca},
    title = {{Hydrogen capture by porphyrins at the TiO$_2$(110) surface}},
    journal = {Physical Chemistry Chemical Physics},
    volume = {17},
    number = {44},
    pages = {30119-30124},
    year = {2015},
    doi = {10.1039/c5cp05437k},
}

@article{2HTPP_ox_TiO2_2,
    author = {Schio, Luca and Forrer, Daniel and Casarin, Maurizio and Goldoni, Andrea and Rogero, Celia and Vittadini, Andrea and Floreano, Luca},
    title = {{On surface chemical reactions of free-base and titanyl porphyrins with r-TiO$_2$(110): a unified picture}},
    journal = {Physical Chemistry Chemical Physics},
    volume = {24},
    number = {21},
    pages = {12719-12744},
    year = {2022},
    doi = {10.1039/d2cp01073a},
}

@article{2HTPP_ox_TiO2_1,
    author = {K\"obl, Julia and Wang, Tao and Wang, Cici and Drost, Martin and Tu, Fan and Xu, Qian and Ju, Huanxin and Wechsler, Daniel and Franke, Matthias and Pan, Haibin and Marbach, Hubertus and Steinr\"uck, Hans-Peter and Zhu, Junfa and Lytken, Ole},
    title = {{Hungry Porphyrins: Protonation and Self-Metalation of Tetraphenylporphyrin on TiO$_2$(110)-1$\times$1}},
    journal = {ChemistrySelect},
    volume = {1},
    number = {19},
    pages = {6103-6105},
    doi = {https://doi.org/10.1002/slct.201601398},
    year = {2016}
}

@article{InPc_1,
    title = {{Spin-state switching of indium-phthalocyanine on Pb(100)}},
    journal = {RSC Advances},
    volume = {14},
    number = {52},
    pages = {38506-38513},
    year = {2024},
    issn = {2046-2069},
    doi = {https://doi.org/10.1039/d4ra07270g},
    author = {Niklas Ide and Arnab Banerjee and Alexander Weismann and Richard Berndt},
}

@article{PC_intro_spin2,
    author = {Fabian Schulz and Mari Ij\"as, Robert Drost and Sampsa K. H\"am\"al\"ainen and Ari Harju and Ari P. Seitsonen and Peter Liljeroth},
    title = {{Many-body transitions in a single molecule visualized by scanning tunnelling microscopy}},
    journal = {Nature Communications},
    volume = {11},
    pages = {229--234},
    year = {2015},
    doi = {10.1038/nphys3212},
}

@article{PC_intro_spin1,
    author = {Aitor Mugarza and Cornelius Krull and Roberto Robles and Sebastian Stepanow and Gustavo Ceballos and Pietro Gambardella},
    title = {{Spin coupling and relaxation inside molecule–metal contacts}},
    journal = {Nature Communications},
    volume = {2},
    pages = {490},
    year = {2011},
    doi = {10.1038/ncomms1497},
}

@article{PC_intro_qubits,
    author = {Urdaniz, Corina and Taherpour, Saba and Yu, Jisoo and Reina-Galvez, Jose and Wolf, Christoph},
    title = {{Transition-Metal Phthalocyanines as Versatile Building Blocks for Molecular Qubits on Surfaces}},
    journal = {The Journal of Physical Chemistry A},
    volume = {129},
    number = {9},
    pages = {2173-2181},
    year = {2025},
    issn = {1089-5639},
    doi = {10.1021/acs.jpca.4c07627},
}

@article{PC_intro_catal1,
    author = {Sorokin, Alexander B.},
    title = {{Phthalocyanine Metal Complexes in Catalysis}},
    journal = {Chemical Reviews},
    volume = {113},
    number = {10},
    pages = {8152-8191},
    year = {2013},
    month = {06},
    issn = {0009-2665},
    doi = {10.1021/cr4000072},
}

@article{Pc_intro_catal2,
    author = {Wu, Yueshen and Liang, Yongye and Wang, Hailiang},
    title = {{Heterogeneous Molecular Catalysts of Metal Phthalocyanines for Electrochemical CO$_2$ Reduction Reactions}},
    journal = {Accounts of Chemical Research},
    volume = {54},
    number = {16},
    pages = {3149-3159},
    year = {2021},
    issn = {0001-4842},
    doi = {10.1021/acs.accounts.1c00200},
}

@article{GOTTFRIED,
    title = {{Surface chemistry of porphyrins and phthalocyanines}},
    journal = {Surface Science Reports},
    volume = {70},
    number = {3},
    pages = {259-379},
    year = {2015},
    issn = {0167-5729},
    doi = {https://doi.org/10.1016/j.surfrep.2015.04.001},
    author = {J. Michael Gottfried},
}

@article{H2PcAu110,
    title = {{Interplay between metal-free phthalocyanine molecules and Au(110) substrates}},
    journal = {Surface Science},
    volume = {606},
    number = {13},
    pages = {1120-1125},
    year = {2012},
    issn = {0039-6028},
    doi = {https://doi.org/10.1016/j.susc.2012.03.010},
    url = {https://www.sciencedirect.com/science/article/pii/S0039602812001057},
    author = {E. Rauls and W.G. Schmidt and T. Pertram and K. Wandelt},
}

@article{H2PcAg111_metalation,
    author = {Bai, Yun and Buchner, Florian and Wendahl, Matthew T. and Kellner, Ina and Bayer, Andreas and Steinrück, Hans-Peter and Marbach, Hubertus and Gottfried, J. Michael},
    title = {{Direct Metalation of a Phthalocyanine Monolayer on Ag(111) with Coadsorbed Iron Atoms}},
    journal = {The Journal of Physical Chemistry C},
    volume = {112},
    number = {15},
    pages = {6087-6092},
    year = {2008},
    doi = {10.1021/jp711122w},
}

@article{H2PcCu111_metalation,
    author = {Chen, Min and R\"ockert, Michael and Xiao, Jie and Drescher, Hans-J\"org and Steinr\"uck, Hans-Peter and Lytken, Ole and Gottfried, J. Michael},
    title = {{Coordination Reactions and Layer Exchange Processes at a Buried Metal–Organic Interface}},
    journal = {The Journal of Physical Chemistry C},
    volume = {118},
    number = {16},
    pages = {8501-8507},
    year = {2014},
    month = {03},
    doi = {10.1021/jp5019235},
}

@article{H2PcAg110_metalation,
    author = {Smykalla, Lars and Shukrynau, Pavel and Zahn, Dietrich
R. T. and Hietschold, Michael},
    title = {{Self-Metalation of Phthalocyanine Molecules with Silver
Surface Atoms by Adsorption on Ag(110)}},
    journal = {The Journal of Physical Chemistry C},
    volume = {119},
    number = {30},
    pages = {17228-17234},
    year = {2015},
    month = {07},
    doi = {10.1021/acs.jpcc.5b04977},
}

@article{MPc_Cu111,
    author = {Pengcheng Chen and Dingxin Fan and Annabella Selloni and Emily A. Carter and Craig B. Arnold and Yunlong Zhang and Adam S. Gross and James R. Chelikowsky and Nan Yao},
    title = {{Observation of electron orbital signatures of single atoms within metal-phthalocyanines using atomic force microscopy}},
    journal = {Nature Communications},
    volume = {14},
    pages = {1460},
    year = {2023},
    doi = {10.1038/s41467-023-37023-9},
}

@article{H2Pc_tauto4,
    author = {K\"{u}gel, Jens and Leisegang, Markus and Bode, Matthias},
    title = {{Imprinting Directionality into Proton Transfer Reactions of an Achiral Molecule}},
    journal = {ACS Nano},
    volume = {12},
    number = {8},
    pages = {8733--8738},
    year = {2018},
    month = {08},
    doi = {10.1021/acsnano.8b04868},
}

@article{H2Pc_tauto3,
    author = {Anna Ros\l{}awska and Katharina Kaiser and Michelangelo Romeo and Elo\"{\i}se Devaux and Fabrice Scheurer and St\'{e}phane Berciaud and Tom\'{a}\v{s} Neuman and Guillaume Schull},
    title = {{Submolecular-scale control of phototautomerization}},
    journal = {Nat. Nanotechnol.},
    volume = {19},
    pages = {738--743},
    year = {2024},
    doi = {10.1038/s41565-024-01622-4},
}

@article{H2Pc_tauto2,
    author = {Benjamin Doppagne and Tom\'{a}\v{s} Neuman and Ruben Soria-Martinez and Luis E. Parra L\'opez and Herv\'e Bulou and Michelangelo Romeo and St\'{e}phane Berciaud and Fabrice Scheurer and Javier Aizpurua and Guillaume Schull},
    title = {{Single-molecule tautomerization tracking through space- and time-resolved fluorescence spectroscopy}},
    journal = {Nat. Nanotechnol.},
    volume = {15},
    pages = {207--211},
    year = {2020},
    doi = {10.1038/s41565-019-0620-x},
}

@article{H2Pc_tauto1,
    author = {Peter Liljeroth  and Jascha Repp  and Gerhard Meyer},
    title = {{Current-Induced Hydrogen Tautomerization and Conductance Switching of Naphthalocyanine Molecules}},
    journal = {Science},
    volume = {317},
    number = {5842},
    pages = {1203--1206},
    year = {2007},
    doi = {10.1126/science.1144366},
}

@article{H2PcCu110-O,
    title = {{H2Pc and pentacene on Cu(110)-(2$\times$1)O: A combined STM and nc-AFM study}},
    journal = {Surface Science},
    volume = {696},
    pages = {121590},
    year = {2020},
    issn = {0039-6028},
    doi = {10.1016/j.susc.2020.121590},
    author = {Angel Garlant and Bret Maughan and Percy Zahl and Oliver L.A. Monti},
}

@article{H2PcAg100,
    author = {Kami\'{n}ski, Wojciech and Antczak, Gra\.{z}yna and Morgenstern, Karina},
    title = {{Bistable H2Pc Molecular Conductance Switch
on Ag(100)}},
    journal = {The Journal of Physical Chemistry C},
    volume = {126},
    number = {39},
    pages = {16767--16776},
    year = {2022},
    month = {09},
    issn = {1932-7447},
    doi = {10.1021/acs.jpcc.2c03485},
}

@article{H2PcAu111,
    title = {{Metal-free Phthalocyanine (H2Pc) Molecule Adsorbed on the Au(111) Surface: Formation of a Wide Domain Along a Single Lattice Direction}},
    journal = {Sci. Technol. Adv. Mater.},
    volume = {11},
    pages = {054602},
    year = {2010},
    doi = {https://doi.org/10.1088/1468-6996/11/5/054602},
    author = {Tadahiro Komeda, Hironari Isshiki, Jie Liu},
}

@article{H2PcTiO2,
    title = {{Bonding of metal-free phthalocyanine to TiO2(110) single crystal}},
    journal = {Solar Energy Materials and Solar Cells},
    volume = {90},
    number = {20},
    pages = {3602-3613},
    year = {2006},
    issn = {0927-0248},
    doi = {https://doi.org/10.1016/j.solmat.2006.06.054},
    url = {https://www.sciencedirect.com/science/article/pii/S0927024806003047},
    author = {P. Palmgren and B.R. Priya and N.P.P. Niraj and M. G\"othelid},
}

@article{InOxlattice,
    author = {M. Marezio},
    title = {{Refinement of the Crystal Structure of In$_2$O$_3$ at two Wavelengths}},
    journal = {Acta Cryst.},
    volume = {20},
    pages = {723--728},
    doi = {10.1107/S0365110X66001749},
    year = {1966}
}

@article{AgostonAlbe,
  title = {{Thermodynamic stability, stoichiometry, and electronic structure of bcc-In${}_{2}$O${}_{3}$ surfaces}},
  author = {Agoston, Peter and Albe, Karsten},
  journal = {Phys. Rev. B},
  volume = {84},
  issue = {4},
  pages = {045311},
  year = {2011},
  month = {Jul},
  publisher = {American Physical Society},
  doi = {10.1103/PhysRevB.84.045311},
  url = {https://link.aps.org/doi/10.1103/PhysRevB.84.045311}
}

@article{Walsh,
  title = {{Nature of the Band Gap of ${\mathrm{In}}_{2}{\mathrm{O}}_{3}$ Revealed by First-Principles Calculations and X-Ray Spectroscopy}},
  author = {Walsh, A. and Da Silva, J. L. F. and Wei, S.-H. and K\"orber, C. and Klein, A. and Piper, L. F. J. and DeMasi, A. and Smith, K. E. and Panaccione, G. and Torelli, P. and Payne, D. J. and Bourlange, A. and Egdell, R. G.},
  journal = {Phys. Rev. Lett.},
  volume = {100},
  issue = {16},
  pages = {167402},
  year = {2008},
  month = {Apr},
  publisher = {American Physical Society},
  doi = {10.1103/PhysRevLett.100.167402},
}

@article{ITO_organics_interface,
    author = {Armstrong, Neal R. and Veneman, P. Alex and Ratcliff, Erin and Placencia, Diogenes and Brumbach, Michael},
    title = {{Oxide Contacts in Organic Photovoltaics: Characterization and Control of Near-Surface Composition in Indium-Tin Oxide (ITO) Electrodes}},
    journal = {Accounts of Chemical Research},
    volume = {42},
    number = {11},
    pages = {1748--1757},
    year = {2009},
    doi = {10.1021/ar900096f},
}

@article{ITO_3,
    title = {{Properties of ITO thin films deposited by RF magnetron sputtering at elevated substrate temperature}},
    journal = {Materials Science and Engineering: B},
    volume = {77},
    number = {1},
    pages = {110--114},
    year = {2000},
    doi = {https://doi.org/10.1016/S0921-5107(00)00477-3},
    url = {https://www.sciencedirect.com/science/article/pii/S0921510700004773},
    author = {E. Terzini and P. Thilakan and C. Minarini},
}

@article{ITO_2,
    title = {{Rapid annealing obtained ITO films with both extremely low infrared emissivity and high visible light transmission for energy-efficient window applications}},
    journal = {Ceramics International},
    volume = {51},
    number = {3},
    pages = {3163--3169},
    year = {2025},
    doi = {https://doi.org/10.1016/j.ceramint.2024.11.291},
    author = {Q. Chen and T. Gong and W. Chen and F. Fang and Y. Feng and S. Chen and D. Liu and T. Liu},
}

@article{CuPcCoPc_STM,
    author = {Hipps, K. W. and Lu, Xing and Wang, X. D. and Mazur, Ursula},
    title = {Metal d-Orbital Occupation-Dependent Images in the Scanning Tunneling Microscopy of Metal Phthalocyanines},
    journal = {The Journal of Physical Chemistry},
    volume = {100},
    number = {27},
    pages = {11207-11210},
    year = {1996},
    doi = {10.1021/jp960422o},
}

@article{Tasker1979,
	doi = {10.1088/0022-3719/12/22/036},
	url = {https://doi.org/10.1088/0022-3719/12/22/036},
	year = {1979},
	month = {nov},
	publisher = {},
	volume = {12},
	number = {22},
	pages = {4977},
	author = {P. W. Tasker},
	title = {{The stability of ionic crystal surfaces}},
	journal = {Journal of Physics C: Solid State Physics},
}

@article{MWCuPc2025,
    author = {Blatnik, M. A. and Calcinelli, F. and Jeindl, A. and Eder, M. and Schmid, M. and \v{C}echal, J. and Diebold, U. and Jacobson, P. and Hofmann, O. T. and Wagner, M.},
    title = {{Molecular Arrangements in the First Monolayer of Cu-Phthalocyanine on In$_2$O$_3$(111)}},
    journal = {Journal of Materials Chemistry C},
    volume = {13},
    number = {00},
    pages = {17650--17661},
    year = {2025},
    doi = {10.1039/D5TC01394A},
}

@article{MW2022CoPc,
    title = {{Adsorption configurations of Co-phthalocyanine on In$_2$O$_3$(111)}},
    author = {M. Wagner and F. Calcinelli and A. Jeindl and M. Schmid and O. T. Hofmann and U. Diebold},
    journal = {Surface Science},
    volume = {722},
    pages = {122065},
    year = {2022},
    issn = {0039--6028},
    doi = {10.1016/j.susc.2022.122065},
    url = {https://www.sciencedirect.com/science/article/pii/S0039602822000504}
}

@article{MW2021Nature,
  author = {Wagner, M. and Meyer, B. and Setvin, M. and Schmid, M. and Diebold, U.},
  title = {{Direct assessment of the acidity of individual surface hydroxyls}},
  doi = {10.1038/s41586-021-03432-3},
  url = {https://doi.org/10.1038/s41586-021-03432-3},
  year = {2021},
  publisher = {Springer Science and Business Media {LLC}},
  volume = {592},
  number = {7856},
  pages = {722--725},
  journal = {Nature}
}

@article{Hagleitner,
  title = {{Bulk and surface characterization of In$_2$O$_3$(001) single crystals}},
  author = {Hagleitner, D. R. and Menhart, M. and Jacobson, P. and Blomberg, S. and Schulte, K. and Lundgren, E. and Kubicek, M. and Fleig, J. and Kubel, F. and Puls, C. and Limbeck, A. and Hutter, H- and Boatner, L. A. and Schmid, M. and Diebold, U.},
  journal = {Phys. Rev. B},
  volume = {85},
  issue = {11},
  pages = {115441},
  numpages = {11},
  year = {2012},
  month = {Mar},
  publisher = {American Physical Society},
  doi = {10.1103/PhysRevB.85.115441},
  url = {https://link.aps.org/doi/10.1103/PhysRevB.85.115441}
}

@article{MWFranceschi2019,
  title = {Growth of {In$_2$O$_3$(111)} thin films with optimized surfaces},
  author = {Franceschi, G. and Wagner, M. and Hofinger, J. and Kraj\ifmmode \check{n}\else \v{n}\fi{}\'ak, T. and Schmid, M. and Diebold, U. and Riva, M.},
  journal = {Phys. Rev. Mater.},
  volume = {3},
  issue = {10},
  pages = {103403},
  numpages = {10},
  year = {2019},
  month = {Oct},
  publisher = {American Physical Society},
  doi = {10.1103/PhysRevMaterials.3.103403},
  url = {https://link.aps.org/doi/10.1103/PhysRevMaterials.3.103403}
}

@article{MW2017water,
    author = {Wagner, M. and Lackner, P. and Seiler, S. and Brunsch, A. and Bliem, R. and Gerhold, S. and Wang, Z. and Osiecki, J. and Schulte, K. and Boatner, L. A.\ and Schmid, M. and Meyer, B. and Diebold, U.},
    title = {Resolving the Structure of a Well-Ordered Hydroxyl Overlayer on {In$_2$O$_3$(111)}: Nanomanipulation and Theory},
    journal = {ACS Nano},
    volume = {11},
    number = {11},
    pages = {11531-11541},
    year = {2017},
    doi = {10.1021/acsnano.7b06387},
}

@article{MW2014,
    author = {Wagner, M. and Seiler, S. and Meyer, B. and Boatner, L. A.\ and Schmid, M. and Diebold, U.},
    title = {Reducing the {In$_2$O$_3$(111)} Surface Results in Ordered Indium Adatoms},
    journal = {Advanced Materials Interfaces},
    volume = {1},
    number = {8},
    pages = {1400289},
    doi = {10.1002/admi.201400289},
    year = {2014}
}

\clearpage
\setstretch{1.25}
\setcounter{page}{1}
\begin{center}
\textbf{\LARGE Supporting Information}\\[1cm]

\textbf{\Large The absence of a central metal ion destabilizes phthalocyanine on \ch{In2O3}(111)}\\[0.5cm]

Viktoria Waidbacher$^1$, Sarah Tobisch$^1$, Faith J.\ Lewis$^1$, Moritz M.J.\ Eder$^1$, Michael Schmid$^1$, Gareth Parkinson$^1$, Ulrike Diebold$^1$, Margareta Wagner$^{1*}$ 
\\[0.5cm]

$^1$Institute of Applied Physics, TU Wien, 1040 Vienna, Austria\\
$^*$Corresponding author: wagner@iap.tuwien.ac.at\\[1cm]
\end{center}

\begin{minipage}{14.6cm}
\def\tabcolsep{2pt}
\def\arraystretch{1.5}
\begin{tabular}{rl@{\qquad}r}
  \multicolumn{3}{l}{\textbf{Additional Experimental Data}}\\
  1) & Overview STM images of H$_2$Pc on \ch{In2O3}(111) at various coverages & 2\\
  2) & H$_2$Pc on \ch{In2O3}(111) imaged at room temperature & 3\\
  3) & Loss of symmetry in H$_2$Pc adsorbed on \ch{In2O3}(111) & 3\\
  4) & Evaluation of the H$_2$Pc adsorption site & 4\\
  5) & Electronic band gap of \ch{In2O3}(111) & 6\\
  6) & Heating H$_2$Pc on \ch{In2O3}(111) to 50\,$^\circ$C & 6\\
\end{tabular}
\end{minipage}
\\[1cm]
Note: All STM images were acquired at $\approx$5\,K, unless otherwise stated.

\setcounter{section}{0}
\section{Overview STM images of H$_2$Pc on \ch{In2O3}(111) at various coverages}

\begin{figure}[H]
    \centering
    \includegraphics[width=0.8\linewidth]{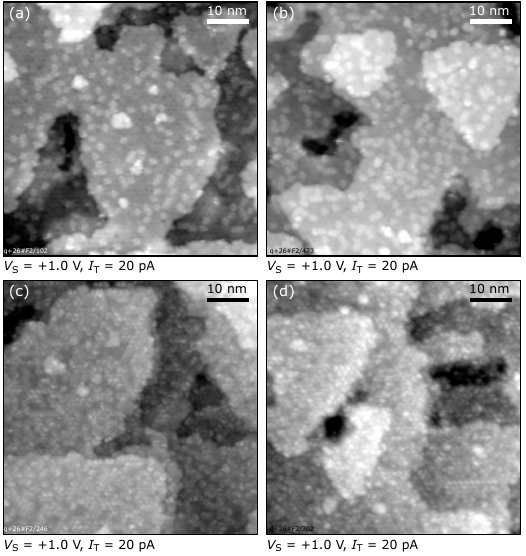} 
    \caption{Overview STM images of (a) 0.2\,ML, (b) 0.4\,ML, (c) 0.5\,ML, and (d) $>$0.6\,ML H$_2$Pc deposited at room temperature on \ch{In2O3}(111).}
    \label{Sfig:coverage}
\end{figure}

\section{H$_2$Pc on \ch{In2O3}(111) imaged at room temperature}

\begin{figure}[H]
    \centering
    \includegraphics[width=0.8\linewidth]{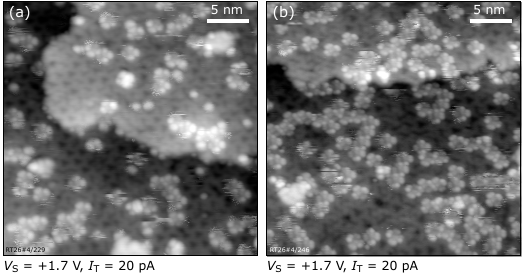} 
    \caption{STM images of H$_2$Pc acquired at 300\,K. Many molecules interact with the tip, visibly as bright and dark horizontal artifacts next to the molecules, but they are otherwise stationary objects.}
    \label{Sfig:rtSTM}
\end{figure}

\section{Loss of symmetry in H$_2$Pc adsorbed on \ch{In2O3}(111)}

\begin{figure}[H]
    \centering
    \includegraphics[width=0.8\linewidth]{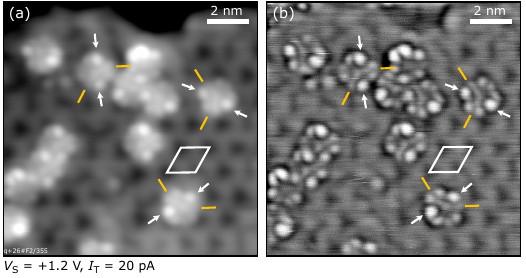} 
    \caption{(a) Empty-states STM image of H$_2$Pc adsorbed on \ch{In2O3}(111), and (b), the same image after high-pass filtering displaying the lack of symmetry of the H$_2$Pc molecules.}
    \label{Sfig:hfiltered}
\end{figure}

Combining two systems that do not share any symmetry operation such as H$_2$Pc on \ch{In2O3}(111), leads to the loss of symmetry. Figure~S\ref{Sfig:hfiltered} shows an STM image acquired at the LUMO energy and after applying a high-pass filter to enhance weak features within the molecules. In addition to the bright lobes on opposite isoindole units (white arrows; resembling mirror symmetry of the molecule) the influence of the threefold symmetric substrate is visible as bright feature connecting two lobes with the center of the molecule (orange lines).

\section{Evaluation of the H$_2$Pc adsorption site}

\begin{figure}[H]
    \centering
    \includegraphics[width=0.8\linewidth]{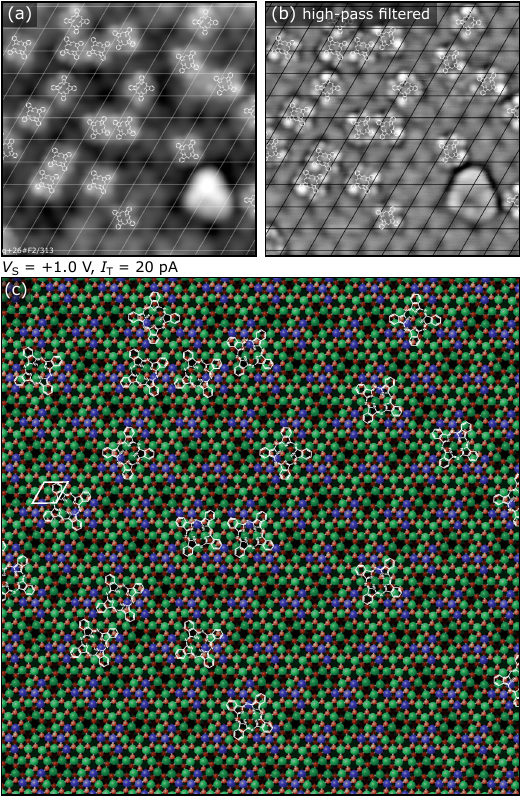} 
    \caption{Identification of the adsorption site of the H$_2$Pc molecules by utilizing the known correlation between the STM contrast and the atomic structure of the \ch{In2O3}(111) surface~\cite{MW2014}. (a) STM image with (1$\times$1) grid aligned to the surface features and H$_2$Pc superimposed; (b) high-pass filtered image.}
    \label{Sfig:site}
\end{figure}

To determine the adsorption site of H$_2$Pc with high accuracy but without atomic resolution of the surface in the presence of the molecules, the following procedure was used: (1) The correlation of the pattern observed in empty-states STM images of the bare surface (center of dark triangles as symmetry point B) to the atomic structure is known from simultaneous constant-height STM/AFM imaging, see Extended Data in Ref.~\cite{MW2021Nature}, and DFT calculations~\cite{MW2014}. (2) To identify the adsorption site of the H$_2$Pc molecules, this correlation is utilized to undistort and align the STM images with the known atomic structure of the substrate, as demonstrated in Figure~S\ref{Sfig:site}. The accuracy of this approach is sufficient to locate the `center' of the H$_2$Pc molecule close to the oxygen anion O($\gamma$), slightly towards the In(5c) atom next to it, and to determine the rotational orientation of the molecule, as discussed in the manuscript. The position of the H atoms at the (pyrrolic) N atoms cannot be extracted from this data and was arbitrarily chosen in this evaluation.

\section{Fundamental band gap of \ch{In2O3}(111)}

\begin{figure}[H]
    \centering
    \includegraphics[width=0.6\linewidth]{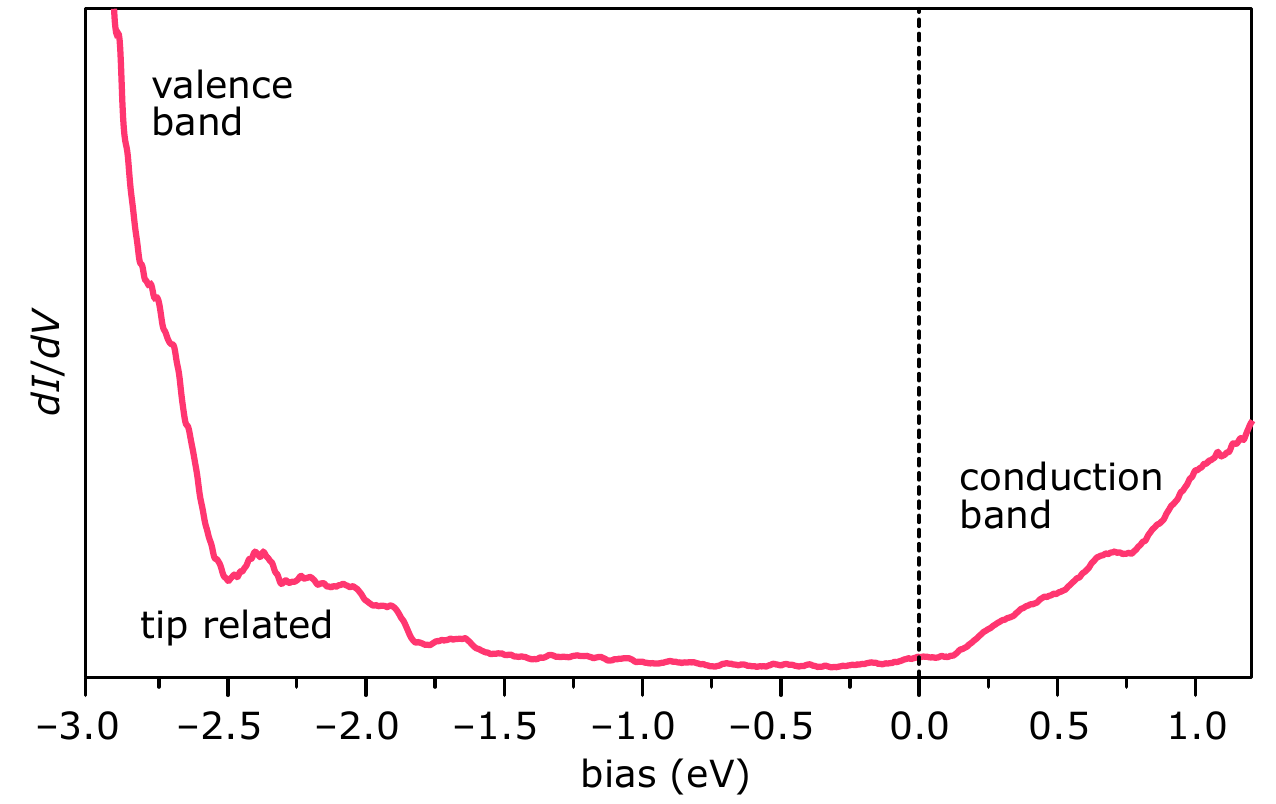} 
    \caption{Local density of states (DOS) of the \ch{In2O3}(111) surface measured by STS with a metallic tip using a lock-in amplifier at 113\,Hz and 1\,mVpp modulation voltage. Note that non-reproducible tip-related states are visible in the band gap close to the valence band. STM imaging is usually acquired by tunneling into empty states.}
    \label{Sfig:bandgap}
\end{figure}

\ch{In2O3} is an n-type semiconductor with a band gap of $\approx$2.9\,eV~\cite{Walsh}. The valence band of \ch{In2O3} consists mainly of O~2p and In~4d states and the conduction band is dominated by In~5s states~\cite{InOx_DFT_bandstructure}. Figure~S\ref{Sfig:bandgap} displays a scanning tunneling spectroscopy (STS) spectrum of the bare surface.

\section{Heating H$_2$Pc on \ch{In2O3}(111) to 50\,$^\circ$C}

\begin{figure}[H]
    \centering
    \includegraphics[width=\linewidth]{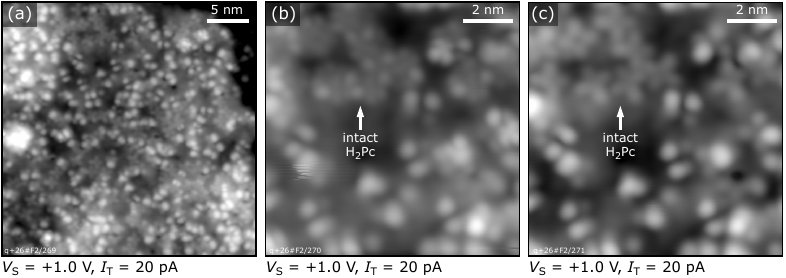} 
    \caption{STM images of H$_2$Pc on \ch{In2O3}(111) after heating  to 50\,$^\circ$C.}
    \label{Sfig:50}
\end{figure}

After heating 0.5\,ML H$_2$Pc on \ch{In2O3}(111) (see Fig~S\ref{Sfig:coverage}c), to 0\,$^\circ$C, approximately 50\% of the molecules have changed in appearance, see Fig.~S\ref{Sfig:50}, and do not resemble intact Pc molecules any more.

\end{document}